\documentclass[prl,reprint,ttisuperscriptaddress,twocolumn,10pt]{revtex4}
\usepackage[dvips]{graphicx}
\usepackage{subfigure}
\usepackage{bm}
\usepackage{color}
\usepackage{amsmath,amssymb}
\begin{document}
\title{
Dynamics of noncollinear antiferromagnetic textures driven by spin current injection
}
\author{Yuta Yamane$^1$, Olena Gomonay$^2$, and Jairo Sinova$^{2,3}$
            }
\affiliation{$^1$Center for Emergent Matter Science (CEMS), RIKEN, Wako, Saitama 351-0198, Japan}
\affiliation{$^2$Institut f\"{u}r Physik, Johannes Gutenberg Universit\"{a}t Mainz,D-55099 Mainz, Germany}
\affiliation{$^3$Institute of Physics Academy of Sciences of the Czech Republic, Cukrovarnicka 10, 162 00 Praha 6, Czech Republic}
\date{\today}
\begin{abstract}
We present a theoretical formalism to address the dynamics of textured, noncolliear antiferromagnets subject to spin current injection.
We derive sine-Gordon type equations of motion for the antiferromagnets, which are applicable to technologically important antiferromagnets such as Mn$_3$Ir and Mn$_3$Sn, and enables an analytical approach to domain wall dynamics in those materials.
We obtain the expression for domain wall velocity, which is estimated to reach $\sim1$ km/s in Mn$_3$Ir by exploiting spin Hall effect with electric current density $\sim10^{11}$ A/m$^2$.
\end{abstract}
\maketitle
Since the prediction of staggered magnetic order\cite{Neel} and its experimental observation in MnO\cite{Shull}, antiferromagnetic (AFM) materials have occupied a central place in the study of magnetism.
The absence of macroscopic magnetization in AFMs, however, indicates that they cannot be effectively manipulated and observed by external magnetic field, which has hindered active applications of AFMs in today's technology.
Research in the emergent field of antiferromagnetic spintronics\cite{Baltz} has shown that electric and spin currents can access AFM dynamics through spin-transfer torques\cite{Nunez,Urazhdin,Xu,Helen2010,Hals,Swaving,Cheng,Yamane,Baldrati} and spin-orbit torques\cite{Jakob,Shiino}.
Similar to ferromagnets, AFMs can also accommodate topologically nontrivial textures such as domain walls (DWs)\cite{Baryakhtar1985,Papanicolau,Bode} and skyrmions\cite{Bogdanov,Barker}, which play crucial roles in spintronics applications, e.g., racetrack memories\cite{Parkin}.
The studies on current-driven dynamics of AFM textures have opened an avenue toward AFM-based technologies.

Recently, AFMs with noncollinear magnetic configurations are generating increasing attention as they exhibit large magneto-transport and thermomagnetic effects;
e.g., anomalous Hall effect\cite{Chen,Kubler,Nakatsuji}, anomalous Nernst effect\cite{Li,Ikhlas} and magneto-optical Kerr effect\cite{Feng,Higo}. 
These phenomena have their origins in the topological character of the electronic band structures, which in turn are associated with the noncollinear magnetism.
To take full advantages of noncollinear AFMs in spintronics applications, it is also important to achieve efficient manipulation of magnetic textures, such as DWs, in those materials.
The studies on current-driven dynamics of AFMs, however, have thus far mostly focused on collinear structures.
Understanding the effects of electric and spin currents in noncollinear AFMs is being a crucial issue in the community\cite{Prakhya,Helen2015,Liu}.

In this paper, we focus on the dynamics of noncollinear AFMs induced by spin current (SC) injection, which may be realized by exploiting spin Hall effect/spin-polarized electric current in an adjacent heavy-metal/ferromagnetic layer. 
We derive sine-Gordon type equations of motion for the AFMs, including effective forces due to SC injection, external magnetic field, and internal dissipation.
Our model can be applied to technologically important triangular AFMs such as Mn$_3$Ir and Mn$_3$Sn.
We then study DW dynamics, where an analytical expression for the DW velocity is derived.

\emph{Model. ---}
We consider an antiferromaget (AFM) composed of three equivalent magnetic sublattices (A, B, and C) with constant saturation magnetization $M_{\rm S}$.
In our coarse-grained model, the classical vector ${\vec m}_A ({\vec r}, t)$ $(|{\vec m}_A ({\vec r}, t)| = 1)$ is a continuous field that represents the magnetization direction in the sublattice A, with similar definitions for ${\vec m}_B ({\vec r}, t )$ and $ {\vec m}_C ({\vec r}, t )$ (Fig.~1).

The magnetic energy density $u$ of the AFM is modeled as follows,
\begin{eqnarray}
  u  &=&   J_0 \sum_{\langle \zeta \eta \rangle} {\vec m}_\zeta \cdot {\vec m}_\eta
             + A_1 \sum_{x_i=x,y,z} \sum_{\zeta=A,B,C}
                       \left(\frac{\partial{\vec m}_\zeta}{\partial x_i}\right)^2  \nonumber \\ &&
              - A_2 \sum_{x_i=x,y,z} \sum_{\langle\zeta\eta\rangle} 
                       \frac{\partial{\vec m}_\zeta}{\partial x_i} \cdot \frac{\partial {\vec m}_\eta}{\partial x_i}  
             + D_0 {\vec e}_z \cdot \sum_{\langle \zeta \eta \rangle} 
                                           {\vec m}_\zeta \times {\vec m}_\eta \nonumber \\ &&
             + u_{\rm ani} - \mu_0 M_{\rm S} {\vec H} \cdot \sum_{\zeta=A,B,C} {\vec m}_\zeta  ,
  \label{u}
\end{eqnarray}
where $J_0$ describes antiferromagnetic exchange coupling between the sublattices, $A_1$ and $A_2$ are the isotropic exchange stiffnesses\cite{Note}, $D_0$ characterizes the homogeneous Dzyaloshinskii-Moriya interaction (DMI), ${\vec H}$ is the external magnetic field, and $\mu_0$ is the vacuum permeability.
The symbol $\langle \zeta \eta \rangle$ indicates the sum over the pairs $(\zeta, \eta) = (A,B)$, $(B,C)$, and $(C,A)$.
For the anisotropy part $u_{\rm ani}$ we assume
\begin{equation}
  u_{\rm ani}  =  - K \sum_{\zeta=A,B,C} \left( {\vec e}_\zeta \cdot {\vec m}_\zeta \right)^2  ,
  \label{u_ani}
\end{equation}
where $K (>0)$ is the anisotropy constant, and the unit vectors ${\vec e}_\zeta$ indicate the easy axes for ${\vec m}_\zeta$ in the $x$-$y$ plane;
${\vec e}_A = ( - {\vec e}_x + \sqrt{3} {\vec e}_y ) / 2$, ${\vec e}_B = - ( {\vec e}_x + \sqrt{3} {\vec e}_y ) / 2 $, and $ {\vec e}_C = {\vec e}_x $.
The magnetic anisotropy of this form applies to triangular AFMs such as the L1$_2$ phase of Mn$_3$Ir\cite{Ulloa} and the hexagonal phase of Mn$_3$Sn\cite{Liu}, with the (1,1,1) plane of their fcc crystals identified as our $x$-$y$ plane.
Although there can also be a smaller out-of-plane anisotropy in realistic materials, Eq.~(\ref{u_ani}) suffices for our present purpose of understanding the fundamental response of triangular AFMs to SC injection.

\begin{figure}
  \centering
  \includegraphics[width=8cm,bb=0 0 436 477]{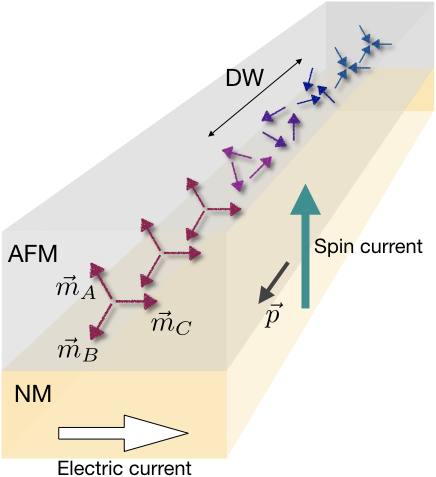}
  \caption{ Scheme of the studied system;
                 bilayer of noncollinear antiferromagnet (AFM) and nonmagnetic (NM) heavy metal, where the spin current with polarization along ${\vec p}$ is injected into the AFM. 
                 Spin current may be created via spin Hall effect (as shown) or by alternative techniques, such as spin-pumping and injection of spin-polarized electric current from a ferromagnetic layer. 
                 Domain wall (DW), connecting the all-in (blue) and all-out (red) domains, is driven into motion by the spin current.
                               }
  \label{fig01}
\end{figure}

The dynamics of ${\vec m}_\zeta$ ($\zeta =A,B,C$) are assumed to obey the coupled Landau-Lifshitz-Gilbert equations;
\begin{equation}
  \frac{\partial{\vec m}_\zeta}{\partial t}  =  - {\vec m}_\zeta \times \gamma {\vec H}_\zeta
                                                                  + \alpha {\vec m}_\zeta \times
                                                                                \frac{\partial{\vec m}_\zeta}{\partial t}
                                                                  - {\vec m}_\zeta \times
                                                                                            \left( {\vec m}_\zeta \times {\vec p} \right)  ,
  \label{llg}
\end{equation}
where $\gamma$ and $\alpha$ are the the gyromagnetic ratio and the Gilbert damping constant, respectively, which are assumed for simplicity to be sublattice independent, and $ {\vec H}_\zeta =  - (\mu_0 M_{\rm S})^{-1} \delta u / \delta {\vec m}_\zeta$ is the effective magnetic field for the sublattice $\zeta$.
The last term in Eq.~(\ref{llg}) is the Slonczewski-Berger spin-transfer torque\cite{Slonczewski} due to SC injection.
The vector ${\vec p}$ represents the value and polarization of the SC, which depend on the way of SC injection, device materials, geometry, etc.
We have assumed that the injected SC transfers the angular momentum equiprobably to each of the sublattices\cite{Helen2015}.

\emph{In-plane triangular approximation. ---}
We here introduce
\begin{eqnarray}
  {\vec n}_1  &=&  \frac{{\vec m}_1 + {\vec m}_2 - 2 {\vec m}_3}{3\sqrt{2}}  ,  \qquad
  {\vec n}_2       =  \frac{- {\vec m}_1 + {\vec m}_2}{\sqrt{6}}  , \\
  {\vec m}  &=&  \frac{{\vec m}_1 + {\vec m}_2 + {\vec m}_3}{3}  .
\end{eqnarray}
Because the AFM exchange coupling responsible for the formation of triangular structure is usually dominant over the other energies, one can safely assume $|{\vec m} ({\vec r}, t)| \ll 1$.
The vectors $ {\vec n}_1 $ and ${\vec n}_2$ are then approximated to be orthogonal to each other and have the fixed length as $ | {\vec n}_1 | \simeq | {\vec n}_2 | \simeq 1/ \sqrt{ 2 } $.
These two vectors can be regarded order parameters of the AFM\cite{Helen2015}, specifying the particular triangular configuration.

We further assume that the in-plane anisotropy is sufficiently large that the triangle is formed in the $x$-$y$ plane with $ | m_\zeta^z | \ll 1$, $\forall\zeta$.
This leads to an approximation where only the $x$ and $y$ components of $ {\vec n}_1 $ and $ {\vec n}_2 $ are nonzero (while ${\vec m}$ can still have a finite $z$ component).
In this case the orientations of ${\vec n}_1$ and ${\vec n}_2$ in the $x$-$y$ plane can be parameterized by a single azimutal angle $\varphi$\cite{Andreev,Helen2015,Ulloa,Liu} as
\begin{eqnarray}
  {\vec n}_1  &=&  \frac{1}{ \sqrt{ 2 } }\left[ \begin{array}{c}  \cos \varphi  \\  \sin \varphi  \\  0  \end{array} \right]  ,  \label{n1} \\
  {\vec n}_2  &=&  {\cal R}_{ \pm \pi / 2 } {\vec n}_1  \equiv  \frac{1}{ \sqrt{ 2 } }\left[ \begin{array}{c}  \cos ( \varphi \pm \pi / 2 )  \\  \sin ( \varphi \pm \pi / 2 )  \\  0   \end{array} \right]  .
\label{n2}
\end{eqnarray}
In Eq.~(\ref{n2}), ${\cal R}_{+\pi/2}$ and ${\cal R}_{-\pi/2}$ select the $+\pi/2$ and $-\pi/2$ rotations of ${\vec n}_2$ against ${\vec n}_1$, respectively, corresponding to the two different chiralities of the triangular structure, defined by ${\rm sgn} \left( {\vec e}_z \cdot {\vec n}_1 \times {\vec n}_2 \right)$;
in Fig.~2, four different triangular configurations are shown as examples.
Which of ${\cal R}_{+\pi/2}$ and ${\cal R}_{-\pi/2}$ should be chosen is dictated by the DMI and magnetic anisotropy.
The DMI favors the ${\cal R}_{+\pi/2}$ (${\cal R}_{-\pi/2}$) chirality if the sign of $D_0$ is negative (positive).
The magnetic anisotropy, on the other hand, can never be fully respected by ${\cal R}_{-\pi/2}$ [Fig.~2.~(c) and (d)], in contrast to ${\cal R}_{+\pi/2}$ where the anisotropy energy is minimized by taking $\varphi=0,\pi$. [Fig.~2.~(a) and (b)]
The ${\cal R}_{-\pi/2}$ chirality is thus favored when the DMI satisfies the condition $2\sqrt{3}D_0 > K$.
As a result of the competition between the anisotropy, exchange coupling and DMI, the ${\cal R}_{-\pi/2}$ triangles carry the weak in-plane ferromagnetic moment ${\vec m}$ [Fig.~2.~(c) and (d)].
Typical materials that host the ${\cal R}_{+\pi/2}$ triangles include the L1$_2$ phase of IrMn$_3$\cite{Sakuma,Szunyogh}, while the ${\cal R}_{-\pi/2}$ configurations are observed in, e.g., the hexagonal phase of Mn$_3$Z (Z $=$ Sn, Ge, Ga)\cite{DZhang}.

It turns out that, for either chirality, the AFM responds to the injected SC in a similar way.
In the following we mostly focus on the ${\cal R}_{+\pi/2}$ case, and later consider configurations with ${\cal R}_{-\pi/2}$.

With the parametrization in Eqs.~(\ref{n1}) and (\ref{n2}), the state of an AFM is described by four variables $(\varphi,{\vec m})$.
By rewriting Eqs.~(\ref{llg}) in terms of $(\varphi,{\vec m})$ and assuming $J_0 \gg |D_0| \gg K$, one obtains, up to the first order of ${\vec m}$, the closed equation of motion for $\varphi$,
\begin{equation}
     c^2 \Box \varphi
       - \frac{3 \omega_E \omega_K}{2} \sin 2\varphi
  =  3 \omega_E p_z  
 - \gamma \frac{\partial H_z}{\partial t}    
   + \omega_\alpha \frac{\partial\varphi}{\partial t} ,
\label{eom}
\end{equation}
and the explicit expression for ${\vec m}$,
\begin{equation}
  {\vec m}  =  \frac{1}{ 3\omega_E} \left( - \frac{\partial \varphi}{\partial t} {\vec e}_z + \gamma {\vec H}  \right)  ,
\label{m}
\end{equation}
for the ${\cal R}_{+\pi/2}$ case.
We have introduced $ \omega_E = \gamma J_0 / \mu_0 M_{\rm S} $, $\omega_K = 2 \gamma K / \mu_0 M_{\rm S}$, $ \omega_\alpha = 3 \alpha \omega_E $, and $\Box = \nabla^2 - (1/c^2) \partial^2/\partial t^2$ with $ c = \sqrt{3 \omega_E \gamma (2A_1+A_2) / \mu_0 M_{\rm S}}$ the group velocity of spin wave.

\begin{figure}
  \centering
  \includegraphics[width=8cm,bb=0 0 738 659]{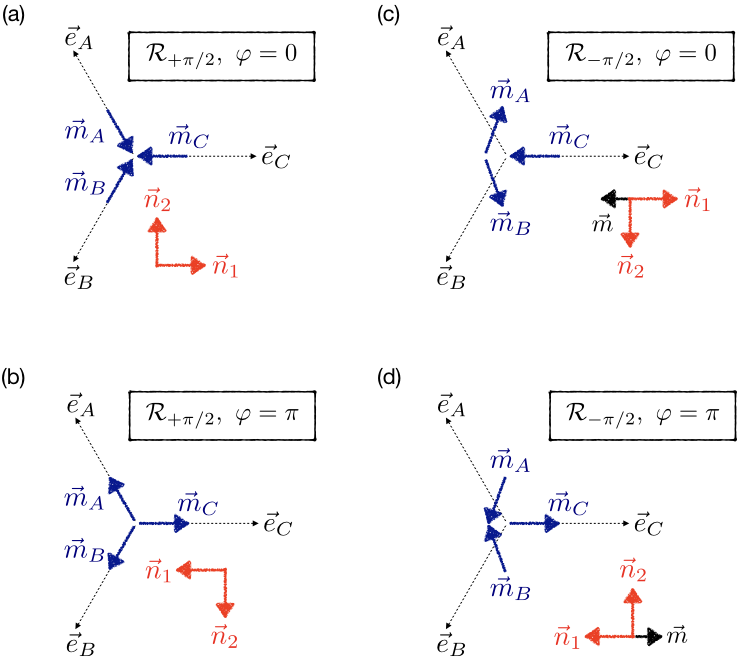}
  \caption{  Triangular magnetic configurations with (a,b) ${\cal R}_{+\pi/2}$ and (c,d) ${\cal R}_{-\pi/2}$ chiralities.
                  They are parametrized by (a,c) $\varphi=0$ and (b,d) $\varphi=\pi$.
                 Dotted lines indicate the easy axes of the magnetic anisotropy.
                }
  \label{fig02}
\end{figure}

Equation~(\ref{m}) shows that $ {\vec m} $ is expressed in terms of $ \varphi $, and vanishes in the absence of magnetic dynamics ($\partial \varphi/\partial t=0$) and external magnetic field.
Equation~(\ref{eom}) is one of our main results.
The rhs of this equation contains the effective forces originating from SC injection, time-varying magnetic field, and internal damping.
In the absence of these forces, Eq.~(\ref{eom}) is reduced to a sine-Gordon equation, consistent with the work in Ref.~\cite{Ulloa}.
In the limit of homogeneous systems ($\nabla\varphi=0$) without external magnetic field, Eq.~(\ref{eom}) then reproduces the result of Ref.~\cite{Helen2015}.
Now, our Eq.~(\ref{eom}) allows one to study inhomogeneous AFM textures under the external driving forces due to SC and magnetic field, and the internal damping.

Notice that only the out-of-plane ($z$) component of the SC polarization and of the magnetic field can induce the dynamics of $\varphi$.
We should also remark that the DMI does not appear in Eqs.~(\ref{eom}) and (\ref{m}).
This is because the DMI energy, which can be written as $3 \sqrt{3} D_0{\vec e}_z \cdot {\vec n}_1\times {\vec n}_2$, is constant within the present approximation, and its contribution to the equations of motion is higher-order.
The DMI plays a crucial role, however, in lifting the degeneracy with respect to the chirality (Fig.~2) as discussed above.

In the following, we discuss the translational motion of a DW driven by SC (setting ${\vec H}=0$).

\emph{Domain wall dynamics. ---}
The doubly-degenerate ground states for the ${\cal R}_{+\pi/2}$ case are given such that the magnetic anisotropy energy is minimized by $\varphi = 0$ and $\pi$, corresponding to the all-in and all-out configurations, respectively [Fig.~2 (a) and (b)].
A DW can be formed as a transition region connecting the two ground states (Fig.~1).
Here let us consider a one-dimensional DW extending along the $z$-axis.
(Due to the isotropic character of the exchange stiffnesses, our conclusions will be independent of the choice of the direction of DW extension, as long as the SC is polarized along the $z$ axis so defined.)
When the rhs of Eq.~(\ref{eom}) is absent, a standard solution $\varphi_{\rm e} (z)$ for a static DW with the boundary condition $\varphi_{\rm e} (z=\pm\infty) = (0,\pm\pi)$ or $(\pm\pi,0)$ (notice that $\varphi$ is defined in $-\pi\leq\varphi\leq\pi$) is obtained as $ \varphi_{\rm e} (z)  =  2 F \tan^{-1} \left[  \exp \left( Q \frac{ z - z_0 }{ \Delta_0 } \right) \right] $;
 $z_0$ is the coordinate of the DW center, $\Delta_0 = \sqrt{(2A_1+A_2)/2K}$ is the DW width parameter, and $(Q,F)=(\pm1,\pm1)$, satisfying $QF = (1/\pi)\int_{-\infty}^\infty dz (\partial\varphi_{\rm e}/\partial z) $, specifies the boundary condition.

To study steady-motion of the DW driven by SC, we employ the following ansatz
\begin{equation}
  \varphi ( z , t )  =  2 F \tan^{ - 1 } \left[  \exp \left( Q \frac{ z - V t }{ \Delta } \right) \right]  ,
\label{dw}
\end{equation}
where $V$ is the velocity of DW center and $\Delta$ is the dynamical width parameter.
By substituting this ansatz into Eq.~(\ref{eom}), multiplying the subsequent equation by $\sin\varphi$, and integrating it along the $z$-axis from $z=-\infty$ to $+\infty$, one finds the relation $V = (FQ\pi\Delta/2\alpha) p_z $.

In the special case where the DW exhibits an inertial motion in the {\it absence} of the rhs of Eq.~(\ref{eom}), the width parameter $\Delta_{\rm in}$ is given by
\begin{equation}
  \Delta_{\rm in}  =  \Delta_0 \sqrt{1 - \frac{V^2}{c^2}}  .
\label{Delta}
\end{equation}
Equation (\ref{Delta}) implies the Lorentz contraction of the DW, stemming from the Lorentz invariance of the sine-Gordon equation.
The rhs of Eq.~(\ref{eom}) may be regarded perturbation, if $3\omega_E p_z$ and $\omega_\alpha |\partial\varphi/\partial t | \sim \omega_\alpha V/\Delta$ are sufficiently small compared to each term in the lhs.
In this perturbative regime one can use the approximation $\Delta = \Delta_{\rm in}$, which leads to
\begin{equation}
 V  =  FQ \frac{ \mu p_z }{ \sqrt{ 1 + ( \mu p_z / c )^2 } }  ,
\label{V2}
\end{equation}
where we have introduced the DW mobility (in the unit of length)
\begin{equation}
  \mu  =  \frac{\pi\Delta_0}{2\alpha}   .
\label{mu}
\end{equation}
Eq.~(\ref{V2}) is one of our central results, revealing important natures of the DW dynamics.
The sign of $V$ is determined by that of $p_z $, i.e., the polarization of the SC, and the factor $FQ$ that characterizes the DW structure.
For $\mu |p_z| / c \ll 1$, the DW velocity depends linearly on $p_z $ as $V \simeq FQ\mu p_z$.
Importantly, $V$ monotonically increases with $|p_z|$, in a similar manner as in collinear AFMs\cite{Shiino}.
Our result thus indicates that the absence of the so-called Walker breakdown\cite{Schryer} is ubiquitous for general AFMs, and a high DW velocity can be achieved by increasing the SC injection.
The previous studies showed that noncollinear AFMs have an advantage over collinear ones in the large magneto-transport effects\cite{Chen,Kubler,Nakatsuji,Feng,Higo}, which provide efficient ways to detect DWs.
Now that Eq.~(\ref{V2}) reveals that the noncollinear AFMs can accommodate DWs moving as fast as in collinear ones, the former are indeed a potential candidate for future spintronics applications.
In Fig.~3, Eq.~(\ref{V2}) is plotted by the solid line as a function of $\mu p_z$ with $(F,Q)=(+1,+1)$.

\begin{figure}
\centering
\includegraphics[width=8cm,bb=0 0 836 639]{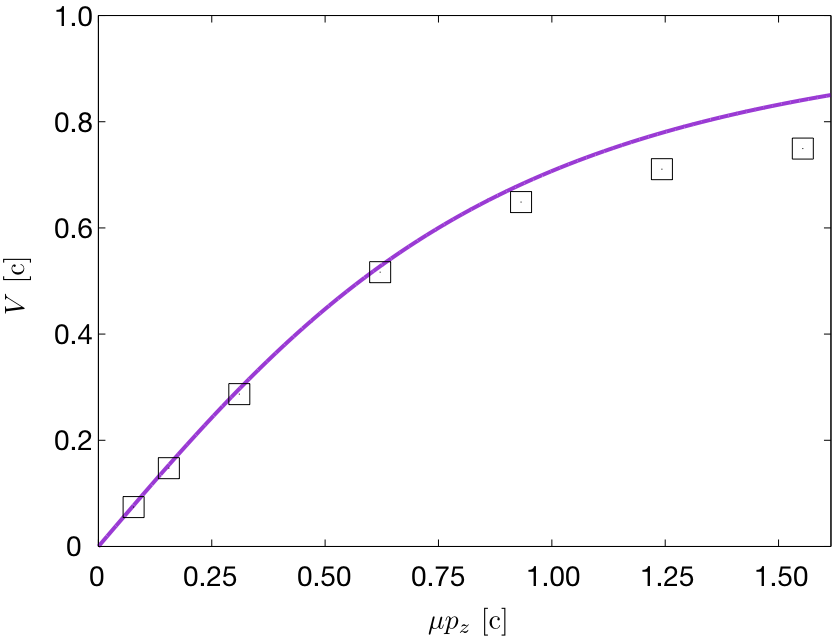}
\caption{ The DW velocity $V$ as a function of SC $\mu p_z$, calculated numerically from Eq.~(\ref{llg}) (open symbols) and analytically from Eq.~(\ref{V2}) (solid line).
Both $V$ and $\mu p_z$ are measured in the unit of $c$ ($\simeq$16 km/s with the present parameter set shown below).
For the simulations we consider a nanowire with dimensions of 1$\times$1 nm$^2\times$20 $\mu$m, dividing it into the unit cells of 1$\times$1$\times$1 nm$^3$.
We employ the periodic boundary condition along the $x$ direction, to mimic a thin (in the $y$ direction), wide (in the $x$ direction) nanowire.
The parameter values are typical for Mn$_3$Ir\cite{Sakuma,Szunyogh}:
$J_0 = 2.4\times10^8 $ J/m$^3$, $A_1=0$ (corresponding to taking into only account the nearest-neighbor coupling in the kagome lattice), $A_2 = 2\times10^{-11}$ J/m, $D_0=-2\times10^7$ J/m$^3$, $K = 3\times10^6$ J/m$^3$, $ \mu_0 M_{\rm S} = 1.63 $ T, $ \gamma = 1.76 \times 10^{ 11 } $ Hz/T,  and $ \alpha = 10^{ - 2 } $.
When the SC is created via spin Hall effect, as in Fig.~1, the electric current density $j_{\rm c}$, corresponding to $\mu p_z = 0.1c$ and $V\simeq4.7$ km/s, is estimated as $j_{\rm c} \simeq 8.5\times10^{11}$ A/m$^2$, using $p_z = (\gamma \hbar / 2eM_{\rm S} ) \theta_{\rm SHE} j_{\rm c} / d$\cite{Slonczewski} with the spin Hall angle $\theta_{\rm SHE}=0.15$ for NM and the sample thickness $d=1$ nm . 
 }
\end{figure}

To check the validity of Eq.~(\ref{V2}), we compute the DW velocity by numerically solving Eq.~(\ref{llg}), as indicated by the open symbols in Fig.~3. 
The results of the simulations and the analytical model agree well in the relatively low current regime.
The discrepancy starts visibly developing as $\mu p_z$ is increased as large as $\sim c$, where the rhs and lhs of Eq.~(\ref{eom}) become comparable (with our present choice of parameters) and the perturbative approach is invalid.
The deviation of the numerical results from Eq.~(\ref{V2}) in the high current regime may be attributed to several factors.
First, the out-of-plane components of the magnetizations grow with $p_z$, thus reducing the accuracy of the in-plane approximation.
Second, the homogeneous SC, represented by the spatial-independent $p_z$ term in Eq.~(\ref{eom}), acts not only within the DW region, but also on each domain.
The SC thus causes the rotation of the domains away from $\varphi=0,\pm\pi$, and the ansatz (\ref{dw}) becomes inappropriate.

\emph{${\cal R}_{-\pi/2}$ chirality. ---}
Lastly, we show that qualitatively similar conclusions are obtained for the ${\cal R}_{-\pi/2}$ case.
The equations of motion are derived with the same line of approximations used in deriving Eqs.~(\ref{eom}) and (\ref{m}).
For the weak ferromagnetic moment ${\vec m}$ one obtains
\begin{equation}
  {\vec m}  =  \frac{1}{3\omega_E} 
                     \left( - \frac{\partial\varphi}{\partial t} {\vec e}_z
                             + \gamma {\vec H}  + \frac{\omega_K}{\sqrt{2}} {\cal M} {\vec n}
                     \right) ,
\label{m2}
\end{equation}
where the external magnetic field ${\vec H}$ is restored, and ${\cal M} {\vec n} = 2^{-1/2} (-\cos\varphi,\sin\varphi,0)$.
Equation (\ref{m2}) differs from Eq.~(\ref{m}) in the third term, which arises from the competition between the magnetic anisotropy, exchange coupling, and DMI, as discussed above.

The equation of motion for $\varphi$, up to the first order of ${\vec m}$, is
\begin{eqnarray}
  c^2 \Box \varphi  &=&  3 \omega_E p_z
                                       - \gamma \frac{\partial H_z}{\partial t} 
                                       + 3 \alpha \omega_E \frac{\partial\varphi}{\partial t}  \nonumber \\ &&
    -  \frac{\omega_K \gamma}{2} \left( H_x \sin\varphi + H_y \cos\varphi \right)  .
\label{eom2}
\end{eqnarray}
There are two major differences between the magnetic dynamics for the ${\cal R}_{+\pi/2}$ [Eq.~(\ref{eom})] and ${\cal R}_{-\pi/2}$[Eq.~(\ref{eom2})] cases;
First, for ${\cal R}_{-\pi/2}$, in-plane magnetic fields can create additional driving forces [the last terms in the rhs of Eq.~(\ref{eom2})], which originates from the direct Zeeman coupling between the weak ferromagnetic moment [the last term in Eq.~(\ref{m2})] and the magnetic field. 
Second, the $\sin2\varphi$ term does not appear in Eq.~(\ref{eom2}), which indicates the absence of effective anisotropy for $\varphi$.
In the ${\cal R}_{-\pi/2}$ case, effective anisotropies arise from higher-order terms of ${\vec m}$\cite{Liu}.
For Mn$_3$Sn, indeed, a small anisotropy $\sim10$ J/m$^3$ of the form of $\cos6\varphi$ has been predicted, which leads to formations of 60$^\circ$ DWs\cite{Liu}.
Although a DW is in general not a 180$^\circ$ wall depending on the symmetry of the effective anisotropy, the SC acts on the DW in essentially the same way as on the 180$^\circ$ walls in the ${\cal R}_{+\pi/2}$ case, since the $p_z$ term is identical in Eqs.~(\ref{eom}) and (\ref{eom2}).
Most of the conclusions on the DW motion derived before thus hold qualitatively, with renormalization $\varphi \rightarrow \frac{n}{2}\varphi$ for a $n$-fold anisotropy with $n$ an even integer.

\emph{Conclusions. ---}
We have derived sine-Gordon type equations of motion for the noncollinear antiferromagnets, with spin current injection, external magnetic field, and dissipative terms included.
We have demonstrated that the injected spin current, when it is polarized perpendicular to the triangular plane, can drive a translational motion of a domain wall.
When the spin current is injected by exploiting the spin Hall effect, the domain wall velocity as high as $\sim1$ km/s can be achieved for typical noncollinear antiferromagnets, with realistic electric current density $\sim10^{11}$ A/m$^2$.
As the spin current injection into noncollinear antiferromagnets remains to be experimentally demonstrated, our findings provide a guideline for devising future experiments.


The authors appreciate Prof. G. Tatara and the members of his group of RIKEN, and Dr. J. Ieda of Japan Atomic Energy Agency,  for fruitful discussions and comments on the manuscript.
This research was supported by Research Fellowship for Young Scientists from Japan Society for the Promotion of Science, the Alexander von Humboldt Foundation, the Transregional Collaborative Research Center (SFB/TRR) 173 SPIN+X, and Grant Agency of the Czech Republic grant No.~14-37427G.
OG also acknowledges the support from DFG (project SHARP 397322108).



\begin{thebibliography}{99}
\bibitem{Note}  The reader may wonder about the minus sign in front of the $A_2$ term.
                        This minus sign ensures the AFM coupling among the different sublattice moments, as well as the positiveness of the spin wave velocity $c$ defined later.
\end{thebibliography}
\end{document}